# Quasi-parallel X-ray microbeam obtained using a parabolic monocapillary X-ray lens with an embedded square-shaped lead occluder

PENG ZHOU,[1,2,3] ZHUXUAN DUO,[1,2,3] ZHIGUO LIU,[1,2,3]

TIANXI SUN,[1,2,3] YUDE LI,[1,2,3] AND SHUANG ZHANG,[1,2,3*]

[1]*Key Laboratory of Beam Technology of the Ministry of Education, College of Nuclear Science and Technology, Beijing Normal University, Beijing 100875, China*
[2]*Applied Optics Beijing Key Laboratory, Department of Physics, Beijing Normal University, Beijing 100875, China*
[3]*Beijing Radiation Center, Beijing 100875, China*
*[*201731220010@mail.bnu.edu.cn](mailto:201731220010@mail.bnu.edu.cn)*

**Abstract:** A parabolic monocapillary X-ray lens (PMXRL) is designed to effectively constrain a laboratory point X-ray source into a parallel beam. A square-shaped lead occluder (SSLO) is used to block direct X-rays in the PMXRL. To design the PMXRL, we use Python to simulate the conic parameter ($p$ = 0.001 mm) of the lens and then use a drawing machine to draw a corresponding lens ($p$ = 0.000939 mm) with a total length of 60.8 mm. We place the SSLO at the lens inlet for optical testing. The results show that the controlled outgoing beam has a divergence of less than 0.4 mrad in the range of 15-45 mm of the lens outlet, which achieves excellent optical performance in X-ray imaging methodology. The design details are reported in this paper.

## 1. Introduction

High-quality quasi-parallel X-rays are important in X-ray analysis, especially in the field of X-ray diffraction analysis (XRD). A polycapillary focusing X-ray lens (PFXRL) combined with a cylindrical capillary collimator (CCC) or a polycapillary parallel X-ray lens (PPXRL) can be applied in X-ray systems because the divergent X-ray beam can be transformed into a quasi-parallel beam to achieve experimental goals[1]. Monocapillary lenses also have the potential to obtain quasi-parallel X-rays because they are effectively applied in the field of X-ray capillary optics. A number of monocapillary lens studies have been reported[2–8]. Drawing towers can be used to draw the hollow glass tube into a target state according to the design requirements[3,9]. Monocapillary lenses are generally divided into parabolic, tapered or ellipsoid capillary lenses according to their shape.

For example, an incident X-ray can be focused to a small size with an ellipsoid capillary lens[9]. The intensity of the X-ray can be enhanced in this small area[5,7,8]. The ellipsoidal monocapillary lens is an ideal focusing optical tool because of its theoretically high focusing performance for a point source. An ellipsoidal monocapillary lens can constrain an X-ray to an angle specified by the profile of the lens inner surface and transport radiation without spreading[5,7]. For the tapered capillary lens, Daniel J. Thiel et al. reported that they used a high-density (5.2 $g/cm^3$) glass material to draw tapered glass monocapillaries, which can be used to obtain micron-width focused beams. However, the beam is susceptible to divergence after exiting the end of the lens because the X-ray in a tapered monocapillary is reflected multiple times in practice[9]. The tested samples should be placed very close to the exit end of the lens to provide a focused spot.

The parabolic monocapillary X-ray lens (PMXRL) is suitable to modulate the divergent point source into an ideal parallel beam or to use it in a reverse path[9]. In 1978, W.T. Welford et al. reported a compound parabolic concentrator[10]. D.X. Balaic et al. reported the focusing of their paraboloidal tapered X-ray focusing capillary, which has an intensity gain of 250±20, into a 6×9 μm pinhole at 8 keV[6]. A. Bjeoumikhov et al. characterized their PMXRL, and the results revealed that the monocapillary could be used over an energy range between 10 and 20 keV and an average full width of half maximum (FWHM) of the focal spot size below 3 μm within that energy range[9]. However, there are very few works on using PMXRLs for obtaining parallel beams.

In the case of the principle of total reflection, the advantages of the PMXRL are that (i) the reflected X-ray can be concentrated to the spatial focus position when the theoretical parallel X-ray passes from the large entrance to the inner wall of the parabolic lens, which is a real geometry optical focus point of the parabolic lens[10]. The theoretical focal spot is infinitely small, which has a significant advantage in microprobe applications and convergence. (ii) The reflection efficiency of the incident X-ray is generally better than 95%. Moreover, the lens is suitable for broad applications in energy detectors because the energy of the incident X-ray is not selected in a parabolic lens, at the detection end of a 3D-XRF tool, for instance. (iii) It is conceivable to obtain an ideal multicolor parallel beam from a point source by adding a beam stop (BS). Therefore, advantage (iii) can be used to obtain quasi-parallel beams in the laboratory. In this work, a parabolic lens is designed with a beam stop to effectively constrain a laboratory X-ray point source into a parallel beam. The details of the design process are reported, and the results show that the controlled outgoing beam reached a divergence of less than 0.4 mrad in the range of 15-45 mm of the lens outlet, which constitutes excellent optical performance in X-ray imaging methodology. The performance of the beam is studied and discussed.

## 2. Methods

*2.1 The raytracing theory of X-rays*

The total reflection principle is satisfied in X-ray optics when the grazing angle of the X-ray is less than the critical angle of total reflection ($\theta_c$). This is an ideal specular reflection strategy to control X-rays with glass optics because the roughness of the glass lens inner surface is usually 0.2-0.5 nm. The parameter $\theta_c$ is related to the incident X-ray energy and the density of the glass material, as shown in Eq. 1:

$$\theta_c = \frac{20.3\sqrt{\rho}}{E_k}, \qquad (1)$$

where $\rho$ is the density of the glass ($g/cm^3$) and $E_k$ is the incident X-ray energy (keV). The real situation can be simulated in a raytracing program according to this principle[11].

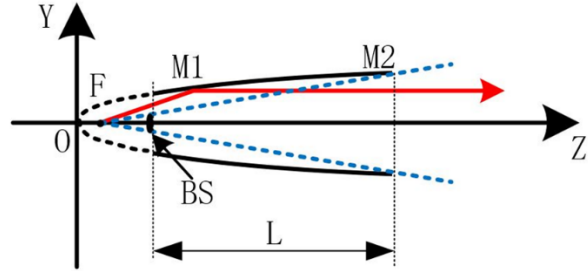

Fig. 1. The profile schematic of a PMXRL with a beam stop.

Figure 1 shows the profile of a PMXRL. The PMXRL is placed in an YOZ cartesian coordinate system with an opening to the right, where L is the actual length of the lens and F is the focal position of the lens. Only reflected X-rays can pass through the lens because a BS is located on the optical path. If an X-ray source is located at position F in the case of Fig. 1, a certain angle of incidence is formed when the X-ray emits from focal point F and enters the PMXRL inner wall. In other words, the X-rays cannot be reflected if the angle of the incident X-ray is more than $\theta_c$. The grazing angle equals $\theta_c$ at position M1, as shown in Fig. 1. This indicates that the rays of all regions from M1 to M2 can be effectively reflected to form an outgoing parallel X-ray. The projection on the Y-axis between M1 and M2 is the bandwidth of the exiting halo.

*2.2 The simulation model of the parabolic monocapillary lens*

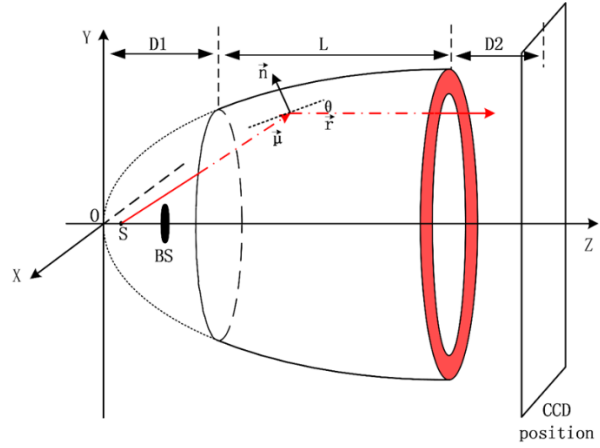

Fig. 2. The 3D simulation model schematic of the PMXRL.

An X-ray passing through the point source is reflected in a parallel manner. We establish a three-dimensional parabolic lens model in the XOYZ coordinate system, as shown in Fig. 2. The parabolic equation in Fig. 2 is written in Eq. 2,

$$x^2 + y^2 = 2pz \qquad (2)$$

Eq. 2 is the internal surface equation of the parabolic lens. The opening is to the right, and the X-ray source (S) is placed at the focus point F. The vertical distance from the coordinate zero point to the incident plane is D1, the length of the lens is L, and the distance from the outlet of the lens to the detector is D2. As shown by the red line in Fig. 2, when the grazing angle is smaller than $\theta_c$, the X-ray satisfies the total reflection principle, and the X-ray is emitted in this simulation. The occlusion BS is added in the optical path according to the simulation needs. However, the addition of the central occluder is extremely complicated in engineering practice and is not convenient for experimental use in actual optical path construction. Therefore, we propose adding a BS in a PMXRL. We aim to apply a square-shaped lead occluder (SSLO) in the PMXRL to block direct rays. The corresponding simulation model is built as follows:

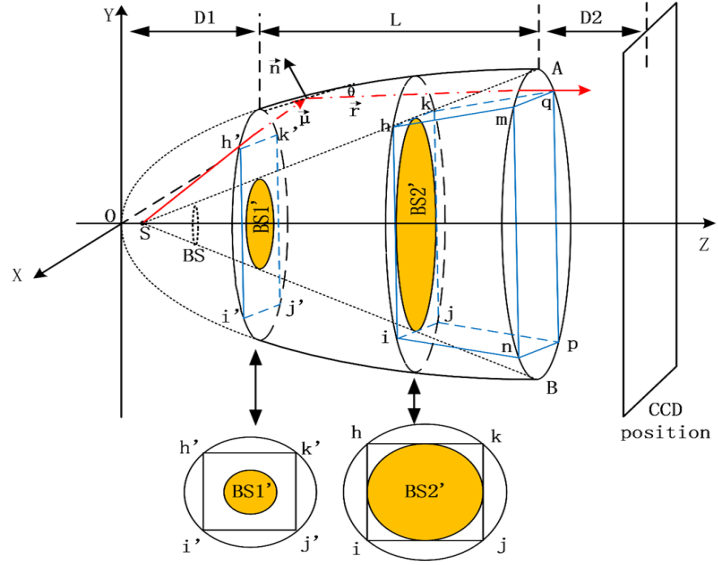

Fig. 3. The simulation model schematic of the PMXRL with the SSLO.

As shown in Fig. 3, in the solid angle ASB, an actual BS can be obtained by adding an occluder at an arbitrary position. The situation of blocking direct X-rays is easy to obtain, such as the positions of BS1' or BS2'. Therefore, we consider that there is a plane named hijk, as shown in Fig. 3, when determining the position of the occluder.

The embedded SSLO in the position is exactly the inscribed square that blocks the round area of solid angle SAB, which can cover all the direct X-rays (BS2') and minimize the loss of reflected X-rays. This is because the side length of the embedded SSLO is larger than the diameter of the circular plane to be shielded (BS1') if we assume the position is located on plane h'i'j'k'. Therefore, there is a suitable plane hijk, as shown in Fig. 3. We incorporate an SSLO named hijk-mnpq embedded in the PMXRL. The simulation results and experimental results are discussed below.

*2.3 Determining the main parameters and drawing the lens*

To ensure that the final PMXRL has good stiffness in the present work[6], we chose a preform tube with an outer diameter of 20.5 mm and an inner diameter of 2 mm. Combining the selected preform tube and the actual drawing configuration, we set the $p$ value to 0.001 in the design process. We set the exit position of the lens to 120 mm; therefore, the solid angle ASB remained unchanged during the simulation to find the best occlusion position. Then, we chose different lens lenses to obtain simulated results. The best occlusion position in each simulation case and the direct or reflect X-ray counts were sequentially recorded. In addition, we used a preform to draw this PMXRL.

## 3. Experimental process and results

*3.1 The simulation results*

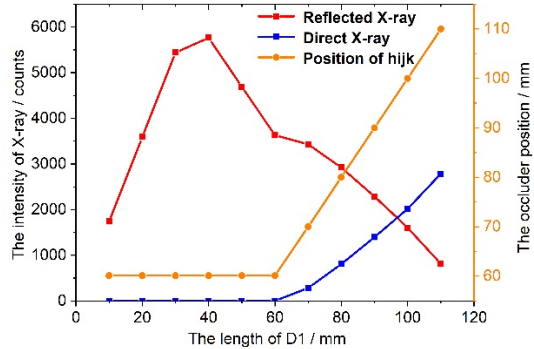

Fig. 4. The simulation results of the PMXRL with SSLO.

The simulation program is written in Python. We set the energy of the X-ray to 8.0 KeV and the source spot to 8 μm. The intensity of the source is set to $10^8$ counts/s. We set $p$ to 0.001 mm for optical simulation so that the shape of the simulated lens did not change. Then, the exit end of the lens is fixed at 120 mm. Therefore, ∠ASB is fixed, which is a prominent feature. Finally, we simulate the interception of the lens at different locations and check whether the SSLO is effectively occluded or overoccluded. In this way, the different inlet ends are simulated. The value of the inlet end is 10 mm with a spacing of 10 mm.

It should be noted that the attenuation of X-rays in the air is ignored during the simulation. To more intuitively compare different simulation parameters, we set the actual number of incident photons per simulation to 10,000. The red line records the reflected X-ray counts, and the blue line records the direct X-ray counts in Fig. 4. The yellow line shows the best occlusion position.

*3.2 The results of drawing*

We preset $p = 0.001$ mm to draw the lens at the beginning of the design. Then, in combination with the tube's outer and inner diameter ratio of 20.5:2, we scanned the outer diameter of the drawn lens and converted it to inner diameter data, as shown in Fig. 5(a)[8]. Fig. 5(b) shows the detailed results of the intercepted lens[12]. The red line shows the fitting result with $(y+a)^2 = 2p(x+b)$. We found that this lens conforms to the parabolic lens type, with $p = 0.000939$ mm[12].

We converted the results into the coordinate system as shown in Fig. 1. Therefore, x1 = 58.4 mm, x2 = 119.2 mm, and the total length is equal to 60.8 mm. We also performed an optical simulation for this situation and found that the results were essentially the same as those described in 3.1. Fig. 5(c) shows a photograph of the intercepted lens. Then, we made a square occluder from lead and placed it at the entrance of the lens for optical experiments.

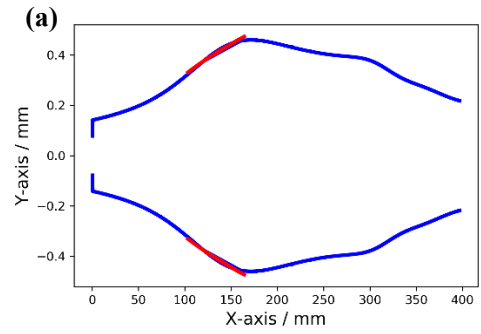

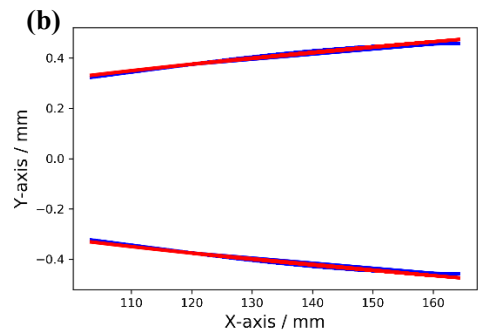

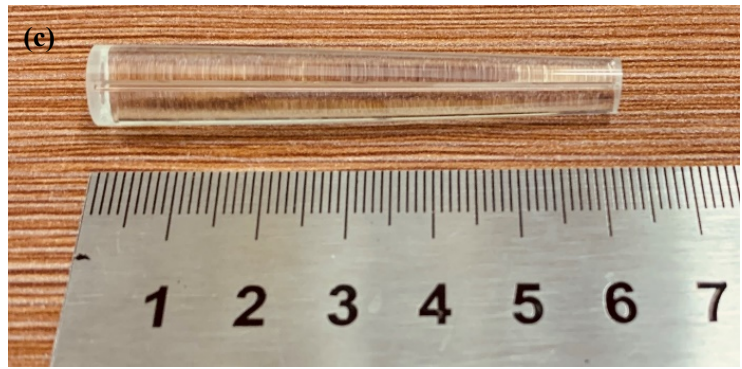

Fig. 5. (a) The inner diameter scan results of the drawn lens (the blue line shows the full lens outline, and the red line shows the intended intercept position). (b) The inner diameter scan results of the intercepted PMXRL (the blue line shows the actual scan result, and the red line shows the standard parabolic equation fit result). (c) Photograph of the intercepted lens corresponding to subfigure (b).

*3.3 X-ray experiment arrangement and results*

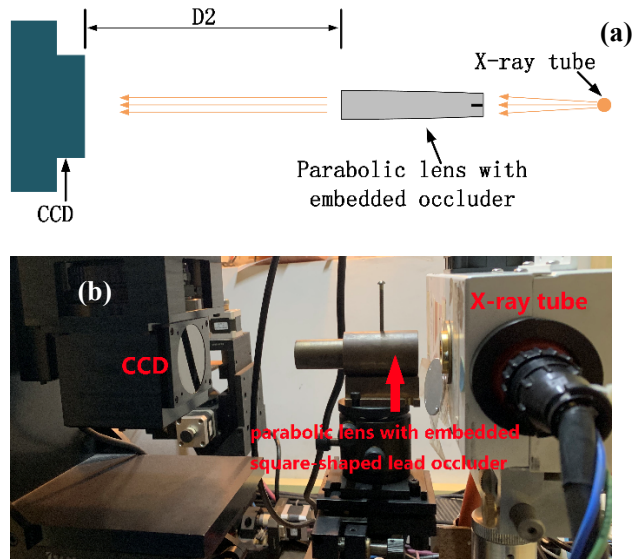

Fig. 6. (a) Schematic diagram of the experimental arrangement. (b) Photograph of the experimental setup.

Figure 6(a) shows a schematic diagram of the experimental arrangement. The PMXRL embedded with an SSLO was tested. Figure 6(b) shows a photograph of the experimental setup. The Cu target X-ray tube [TFX-8100, Trufocus, USA] was operated at 28 KV and 0.14 mA. The size of the X-ray tube focal spot was 8 μm, and the focal spot was placed at the focus position of the PMXRL. A CCD camera [M11427-61, Hamamatsu, Japan] with 10 μm resolution was used to detect the outgoing X-ray spot of the lens. After the optical path was adjusted, we collected the outgoing X-ray spot at different D2 values (15–45 mm) from the lens exit. The exposure time of each group was 10 s. The outgoing X-ray spot photograph is shown in Fig. 7.

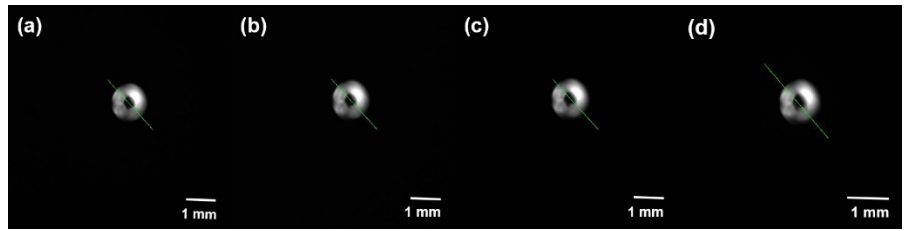

Fig. 7. Photograph of the outgoing X-ray at different D2 positions of the lens exit: (a) 15 mm; (b) 25 mm; (c) 35 mm; (d) 45 mm.

We measured the full width of the spot at different D2 positions when evaluating its divergence because the outgoing X-ray spot is not a normal Gaussian-distributed spot. Taking Fig. 7(c) to obtain the full width of the spot at the 35 mm position as an example, the count profile data were obtained by drawing a green line profile in the CCD camera software [HCimage, Hamamatsu, Japan]. Figure 8 shows the results of the test at 35 mm. Since the origin data and the smoothed data are highly consistent, only the smoothed profile is shown in the results. The red profile is the 9-point Savitzky-Golay smoothed result of the line profile, and the corresponding polynomial order was set to 3. The criteria for selecting this set of parameters are to maintain the shape of the original data profile and to eliminate some fluctuations. The blue curve is the 1st derivative of the red profile, and by identifying its extreme points, the real spot size can be determined better.

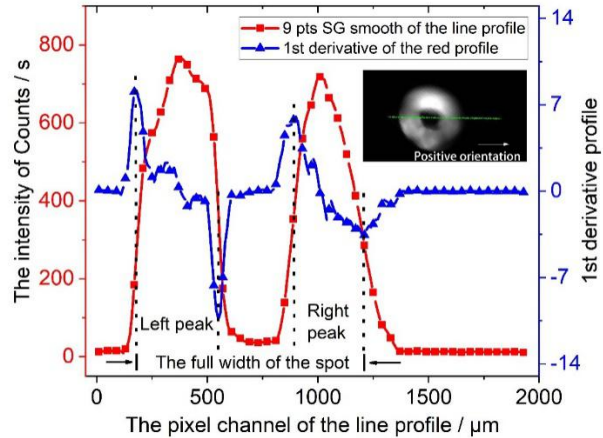

Fig. 8. The results of the Savitzky-Golay smoothed curve of the line profile at 35 mm and its 1st derivative curve. The photo of the real spot in Fig. 8 is obtained by rotating the photo of Fig. 7(c) counterclockwise by 47 degrees; therefore, the positive orientation of the line profile is consistent with the horizontal axis direction of Fig. 8, with the first peak named the left peak and the second peak named the right peak. Due to the picture display, only one point was displayed for every four consecutive points in each curve.

It can be determined that the left peak extended from the 180 μm pixel channel to the 550 μm pixel channel, and the right peak extended from the 900 μm pixel channel to the 1210 μm pixel channel, as shown in Fig. 8. Therefore, the widths of the left peak and the right peak were determined, and the full width of the real optical spot could also be determined. By this method, all the results of different D2 positions could be obtained, and the results are shown in Table 1.

Table 1. The width of the optical spot at different D2 positions.

| D2 position / mm | Left peak width / μm | Right peak width / μm | Full width / μm |
| --- | --- | --- | --- |
| 15 | 360 | 360 | 1040 |
| 20 | 360 | 350 | 1040 |
| 25 | 370 | 340 | 1040 |
| 30 | 360 | 330 | 1030 |
| 35 | 370 | 310 | 1030 |
| 40 | 370 | 320 | 1040 |
| 45 | 370 | 340 | 1040 |

## 4. Discussion

We have proposed a new way to design a PMXRL that successfully adjusts a laboratory point source into a parallel beam by adding an SSLO. First, the assessment of divergence is carried out. From Table 1, it is clear that the full width of the optical spot is not changed in the range of 15–45 mm, which essentially means that the divergence of the outgoing beam is very small. The actual widths of the left and right peaks become unbalanced as the distance corresponding to D2 increases in Tab. 1. The reason for this phenomenon is the deformation of the SSLO. Therefore, we determine that the SSLO corresponding to the left peak does not undergo substantial deformation and that the divergence of the beam can be effectively evaluated by the data of the left peak.

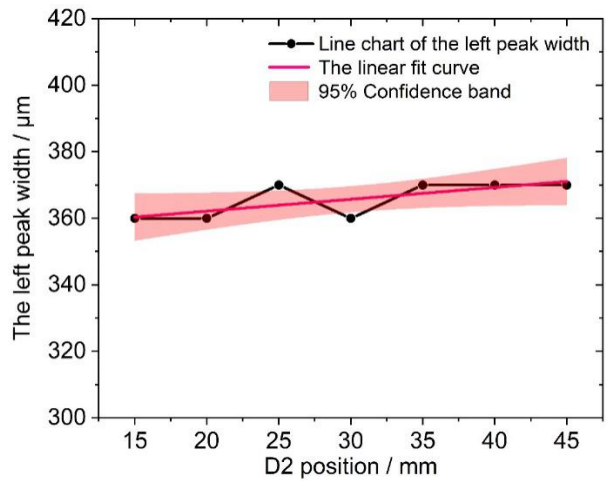

Fig. 9. The divergence of the spot at different D2 positions.

In Fig. 9, the black curve shows the spot data of the left peak at different D2 positions, and the red line is the linear fitting result. Then, the divergence is confirmed to be 0.36 mrad by obtaining parameters of the linear red line. To discuss the result more accurately, the 95% confidence band is confirmed during the fitting process. The 25 mm location data point and the 30 mm point are separated from the 95% confidence band because the resolution of the detector is 10 μm. In general, the divergence of the spot corresponding to the left peak is less than 0.4 mrad. Under the ideal optical path conditions, this result also contributes to the divergence of the outgoing X-ray of the lens.

When the inlet end is in the range of 10-60 mm, as shown in Fig. 4, the occluder can be embedded in the position of 60.1 mm. The complete blocking of direct X-rays in the optical path is satisfied. If the occluder is close to the light source, it blocks not only direct X-rays but also reflected X-rays. Therefore, the direct X-rays are completely blocked when the blocking position is fixed at 60.1 mm. The decrease in reflected X-ray counts at 40-60 mm is due to the increase in the proportion of direct X-rays. The reason why the reflected X-rays at 10-20 mm are significantly lower is based on the fact that they do not satisfy the condition of total reflection.

When the inlet of the PMXRL is located at the position of 60-110 mm, the occluder is unable to effectively block the inlet end of the lens. Therefore, some of the direct X-rays are emitted from the periphery. This kind of situation should be prohibited.

The simulation results ignore the attenuation of X-rays in the air and the actual incident intensity of the X-ray. In each set of simulations, only the ratio of direct and reflected X-rays is considered as well as the actual effective occlusion position. Therefore, the intensity of the reflected X-ray in the experiment would be different from the simulation result in Fig. 4, but the results in Fig. 4 are significant in guiding the drawing and interception positioning of the lens.

We can see that the parabolic monocapillary lens used in this experiment has an excellent contour, as shown in Fig. 5(b). This provides a fundamental guarantee for the favorable experimental results obtained in Fig. 7. The best occlusion position is 60.1 mm according to the simulation, but we intercepted the lens only at 58.4 mm in the engineering process. A certain position error is inevitable when adding an SSLO at the inlet end of the lens. The reflected X-rays are excessively occluded, and the specific influence is reflected in Fig. 7. The deformation is generated at the four corners because the soft lead material is placed inside the PMXRL, which influences the imaging results.

From the results shown of Fig. 7, we did not obtain a central symmetry image. We determined that there is no divergent spot associated with ∠ASB in the outgoing X-ray spot, which indicates that all direct X-rays were blocked. In addition, the occluder excessively blocked the reflected X-rays, and its four corners were deformed, which led to an asymmetrical image. To acquire an ideal outgoing X-ray image, we wrote a 3D simulation program in Python and OpenGL. The obtained simulated outgoing focal spot is shown in Fig. 10.

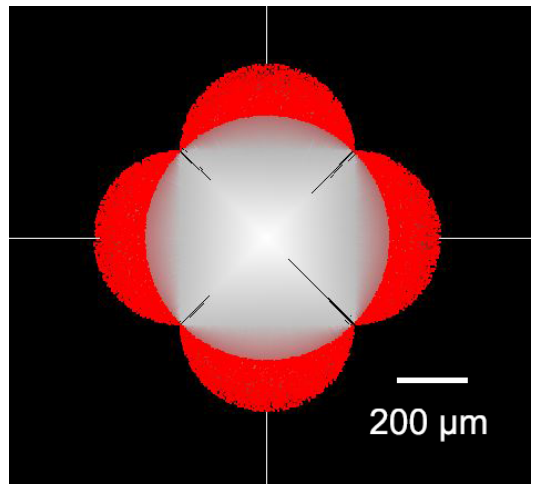

Fig. 10. Simulated outgoing X-ray at the lens exit; the red parts are contributed by the reflected X-ray, the white part is the ray path of the reflected X-ray generated by OpenGL, the square black area is the field of view, and there is no direct X-ray in this field of view.

The four red spots in Fig. 10 are formed by the X-ray reflected from the four sides, and the four spots constitute the ideal outgoing X-ray image. We find that our experimental arrangement initially obtained the expected experimental results, combining the results of Fig. 10 and Fig. 7. It is certain that the direct X-ray in this experiment has been blocked. The occluder was deformed during the insertion process because the lead material was very soft. The divergence of the lens has been evaluated as a collimator. The value of the divergence is 0.36 mrad, and the parallel X-ray emerging from the PMXRL has good parallelism. The lens shows 0.36 mrad divergence in the range of 45 mm. These results are favorable for X-ray analysis methodology at such working distances and divergence. It is possible to add a 20 μm aperture or smaller at the lens exit to ensure the performance of microarea analysis.

We also calculate the intensity of the lens. The X-ray tube in the present work has an 8 μm spot with a 50-degree solid angle. The entrance of the lens was set 58.4 mm from the 8 μm spot. The CCD was 20 mm from the exit surface of the lens. The diameter of the lens entrance is 0.67 mm, and that of its exit is 0.95 mm. The CCD obtains a spot with a 1.55 mm diameter if there is no lens in the optical path, and the intensity of the spot is 0.18 times that of the lens entrance. If the lens is located on the optical path, the CCD captures a spot with a 1.04 mm diameter with the same entrance condition. The intensity is 0.41 times, which can be calculated. Therefore, the intensity generated by the lens 2~3 times that of the case without the lens. Table 2 lists information for four collimators. The results indicate that our PMXRL can be used as an alternative scheme to obtain high-quality quasi-parallel beams in laboratories.

Table 2. The information of different collimators.

| Collimator lists | X-ray energy/keV | Divergence/mrad | Gain/times |
|---|---|---|---|
| Cylindrical Polycapillary[13] | 8.0 | 4.0 | <10 |
| PFXRL with CCC[1] | 8.0 | 4.3 | 120~125 |
| HAMAMATSU J12432-01[14] | 8.0 | 3.49 | 6~7 |
| Our PMXRL | 8.0 | 0.36 | 2~3 |

## 5. Conclusion

In summary, this paper presents a PMXRL with an embedded SSLO. We find that the parameter $p = 0.001$ mm of the parabolic monocapillary lens is suitable for laboratory X-ray sources by taking the optical properties and simulation results.

Through an actual drawing, a lens was fabricated with a total length of 60.8 mm and $p$ of

0.000939 mm. Its optical performance was verified in a laboratory experiment. The results show that the divergence of the outgoing X-ray is less than 0.4 mrad in the range of 15–45 mm. Therefore, the PMXRL with embedded SSLO has the potential for practical application with its extremely low divergence and accommodating working distance.

The limitation of this work is it is not easy to use a lead occluder to obtain a relatively regular imaging spot due to its easy deformation. In future research, hard, high-density materials should be considered for the occluder.

## Acknowledgments

This work was supported by the National Natural Science Foundation of China (Grant No. 11675019, 11875087).